Kyuho Kim†, Kunwoo Park†, Hyungchul Park, Sunkyu Yu\*, Namkyoo Park\*, and Xianji Piao\*

# Programmable photonic unitary circuits for light computing

Programmable photonic unitary circuits

**Abstract:** Unitarity serves as a fundamental concept for characterizing linear and conservative wave phenomena in both classical and quantum systems. Developing platforms that perform unitary operations on light waves in a universal and programmable manner enables the emulation of complex light-matter interactions and the execution of general-purpose functionalities for wave manipulations, photonic computing, and quantum circuits. Recently, numerous approaches to implementing programmable photonic unitary circuits have been proposed and demonstrated, each employing different design strategies that distinctly impact overall device performance. Here, we review foundational design principles and recent achievements in the implementation of programmable photonic unitary circuits, with a particular focus on integrated photonic platforms. We classify the design strategies based on the dimensionality of nontrivial unit operations in their building blocks: lower-dimensional unitary units, such as SU(2) operations, and higher-dimensional ones, such as Fourier transforms. In each category, recent efforts to leverage alternative physical axes, such as the temporal and frequency domains, to address scalability challenges are also reviewed. We discuss the underlying concepts, design procedures, and trade-offs of each design strategy, especially in relation to light-based computing.

## 1 Introduction

One of the fundamental postulates in elaborating traditional quantum theory is to assume an isolated system—one that neither exchanges energy nor matter with its environment [1]. This assumption allows the description of quantum systems using Hermitian Hamiltonians, which accompany mathematically well-established structures: real eigenspectra and orthonormal sets of eigenmodes. According to Stone's theorem [2], the temporal evolutions of quantum states are determined by the corresponding unitary propagators of Hamiltonians. Therefore, characterizing the unitary operations obtained from a quantum system directly corresponds to the complete description of the system as a solvable initial value problem, at least in the conservative regime of quantum physics. In this respect, the core operation principles of quantum computations, such as the gate operations and their circuit assemblies, are also developed under unitarity [3].

As a statistical average of quantum electrodynamics, classical optics often adopts the same assumption by neglecting material, structural, and radiation losses of optical systems [4]. Despite the lack of wavefunctions for photons [5], this radical approximation unveils the correspondence between classical wave equations and the Schrödinger equation, which has been the foundation of the analogy between classical and quantum wave phenomena [6, 7]. As a result, many classical optical phenomena can also be well described through unitary operations with sufficiently high accura-

———

\* Corresponding authors: **Sunkyu Yu**, Intelligent Wave Systems Laboratory, Department of Electrical and Computer Engineering, Seoul National University, Seoul 08826, Korea, E-mail: sunkyu.yu@snu.ac.kr; **Namkyoo Park**, Photonic Systems Laboratory, Department of Electrical and Computer Engineering, Seoul National University, Seoul 08826, Korea, E-mail: nkpark@snu.ac.kr; and **Xianji Piao**, Wave Engineering Laboratory, School of Electrical and Computer Engineering, University of Seoul, Seoul 02504, Korea, E-mail: piao@uos.ac.kr

**Kyuho Kim**, Intelligent Wave Systems Laboratory, Department of Electrical and Computer Engineering, Seoul National University, Seoul 08826, Korea

**Kunwoo Park**, Intelligent Wave Systems Laboratory, Department of Electrical and Computer Engineering, Seoul National University, Seoul 08826, Korea

**Hyungchul Park**, Intelligent Wave Systems Laboratory, Department of Electrical and Computer Engineering, Seoul National University, Seoul 08826, Korea cy. Inversely, achieving unitary operations through tailored optical systems allows not only the modelling of classical and quantum phenomena [8-13], but also the realization of conservative optical functionalities, such as light flow manipulation [14], modal engineering [15, 16], visual perception [17], and optical energy concentration [18].

† **Kyuho Kim** and **Kunwoo Park** contributed equally to this work.

Recently, light-based computing, especially in terms of data-driven [19-22] and quantum [23, 24] computations, has attracted substantial attention to exploit the unique advantages of signal processing using photons—ultrahigh speed, low energy dissipation, and stable quantum states. These next-generation computing systems have sparked a surge of interest in realizing programmable and universal unitary operations in photonic platforms, which are essential ingredients for trainable photonic deep learning and general-purpose quantum computation. Among various optical platforms, such unitary operations are well developed mostly in photonic integrated circuits [25-27] in order to utilize their small footprint, stable light manipulation, and the elementwise modulation essential for reconfigurability. The rapid development of the field in terms of design [28-30] and fabrication [31-33] also supports the high-fidelity realization of universal unitaries for computing applications.

In this review, we investigate and classify various approaches of constructing programmable and universal forms of $N$-dimensional unitary operations in integrated photonics, focusing on their design philosophy, theoretical and numerical assessments, and pros and cons in terms of application devices. In Section 2, we briefly review the universality of unitary operations in photonics and its connection to light-based computing. In Sections 3 and 4, we exhibit the classification of design strategies according to the dimensionality of nontrivial unit operations, discussing the building blocks of each method, their implementation platforms, and the following performances, such as fidelity and scalability. In the last section, we summarize the review and outlook the potential research topics, focusing on remaining challenges in scalability and fidelity.

## 2 Unitary Operations in Photonics

Unitary operations of light waves are observed ubiquitously in various photonic platforms. The most straightforward example is an isolated system composed of optical elements, such as coupled waveguides [8] and resonators [34], photonic crystals [18], and dielectric metamaterials [35] (Fig. 1a). When the interactions with its environment, such as radiative or material loss, amplification, and dynamical changes of system parameters, are negligible within the temporal range of interest, the system can be well modelled with the following Schrödinger-form equation using the Hermitian Hamiltonian $H$:

$$i\frac{d}{d\xi}|\Psi\rangle = H|\Psi\rangle, \tag{1}$$

where $|\Psi\rangle$ is the state of light inside the system, and $\xi$ depicts the evolution variable that can denote a spatial or temporal axis. The evolution of an optical state within the system is determined by the relation, $|\Psi(\xi)\rangle = \exp(-iH\xi)|\Psi(0)\rangle$, where the propagator $\exp(-iH\xi)$ is unitary due to the Hermiticity of $H$ [2]. For example, light flows inside coupled waveguides or resonators are fully described by the unitary operator $\exp(-iH\xi)$ for any forms of the initial state $|\Psi(0)\rangle$. Notably, the platforms manipulating $H$ in a reconfigurable manner have been widely applied to mimic the state evolutions in the corresponding quantum systems [10-13, 36].

On the other hand, unitary operations also play a critical role in open systems. Consider an open system in photonics (Fig. 1b), which undergoes scattering [37] and diffraction [38]. When the system allows for incident and outgoing wave flows, while interactions with other physical domains—again, manifested as material gain and loss, or system dynamics—are negligible, light behaviors around the system are fully described by the scattering matrix:

$$|\Psi_O\rangle = S|\Psi_I\rangle, \tag{2}$$

where $|\Psi_I\rangle$ and $|\Psi_O\rangle$ denote the incident and scattering waves, respectively, of the same dimension. Due to energy conservation, the relationship $\langle\Psi_O|\Psi_O\rangle = \langle\Psi_I|S^\dagger S|\Psi_I\rangle = \langle\Psi_I|\Psi_I\rangle$ holds for arbitrary $|\Psi_I\rangle$, leading to $S^\dagger S = I$. Therefore, the scattering matrices $S$ for open systems, satisfying $S^\dagger S = SS^\dagger = I$, correspond to unitary operations.

The above categorization of isolated and open systems is not a rigorous one but rather one of convenience. For example, when we consider recent interests on exploiting a temporal axis [39-44], the evolution along $\xi = t$ can be considered the scattering process with the scattering matrix $\exp(-iHt)$, where $H$ describes the temporal variation of an optical potential. Similarly, the scattering matrix description with $S$ can be understood as the energy exchange between eigenmodes when the entire space-time domain including the environment is considered an isolated system for light.

Notably, as long as the system preserves energy of light waves, light behaviors in both isolated and open linear systems can be characterized by unitary operations.

Therefore, achieving high-dimensional, reconfigurable, and stable unitary operations of light through tailored photonic systems is a pivotal step toward advanced functionalities using light signals [25, 26]. For example, the recent surge of interest in photonic deep learning accelerators [45] has stimulated the efforts on achieving high-dimensional unitary operations for modelling weight matrices. When applying the singular value decomposition to an arbitrary weight matrix $W$, as $W = UDV^{\dagger}$, where $U$ and $V$ are unitary matrices and $D$ is a diagonal matrix, realizing a universal and reconfigurable form of unitary matrices enables trainable weight matrices for deep learning. The conservative nature of unitary operations has also attracted substantial attention for realizing unitary neural networks [46] to address the gradient explosion or vanishing problem during the learning process [47-49]. Another important application includes unit operations for linear optical quantum computing (LOQC) [23]. Because quantum computation generally exploits isolated systems, elementary gates and their circuit assemblies dictate the use of unitary optical elements. Therefore, the implementation of universal unitary operations is indispensable for reconfigurable and higher-dimensional photonic quantum computing [50-54]. In terms of classical light devices, reconfigurable diffractive and scattering platforms with unitary natures are also crucial for free-space data-driven computation [55, 56], and high-quality optical imaging, detection, and beam manipulations [57-60] from energy conservation. In these classical, quantum, and data-driven applications, we emphasize that the dimension of the obtained unitary matrix operations determines the ultimate information capacity supported by the system: the number of neurons in deep learning accelerators, the dimension of quantum gates for qudit operations, and the number of accessible channels in classical light devices.

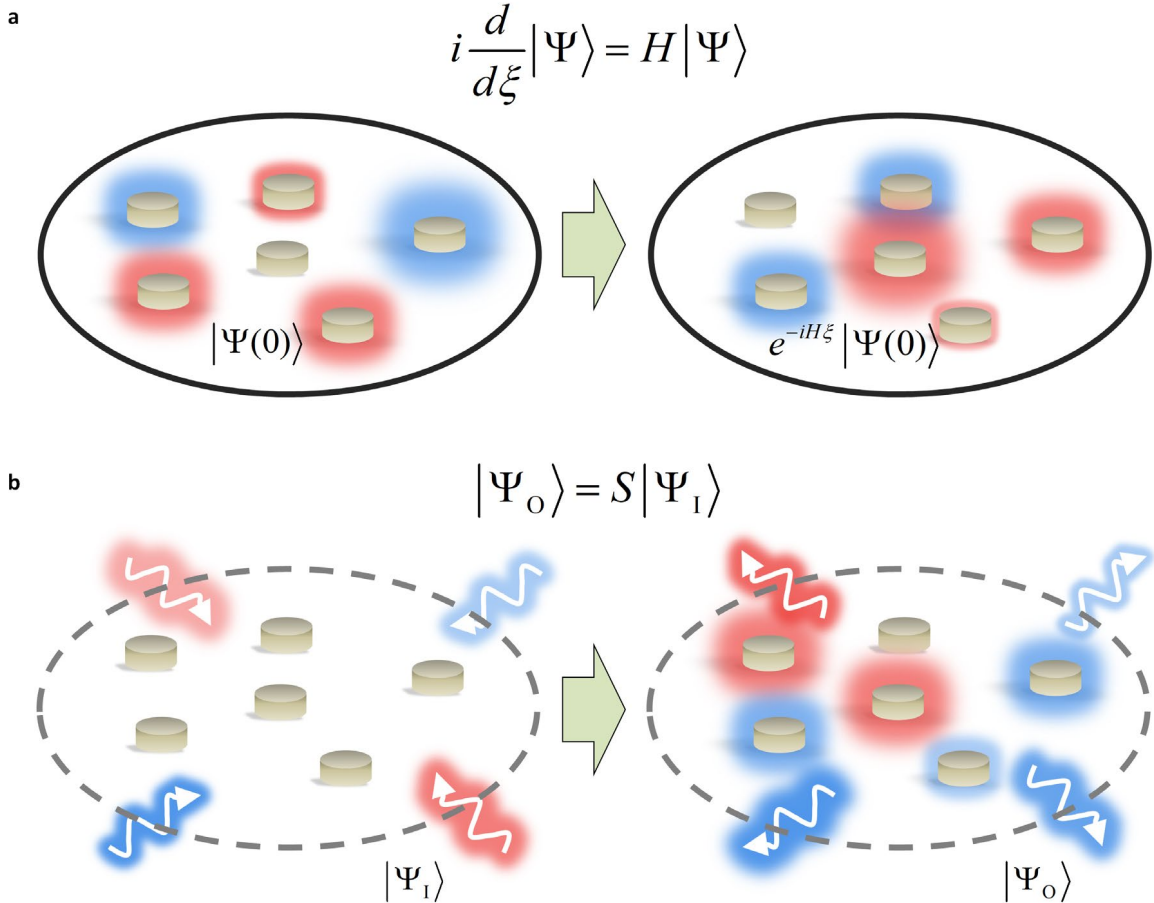

**Figure 1. Unitary operations in photonics.** (a) Schematic for unitary operations in isolated systems. The initial state $|\Psi(0)\rangle$ is evolved to be $|\Psi(\xi)\rangle = \exp(-iH\xi)|\Psi(0)\rangle$, which is governed by the unitary propagator $\exp(-iH\xi)$ from the Hermitian Hamiltonian $H$. (b) Schematic for unitary operations in open systems. The input state $|\Psi_I\rangle$ evolves to $|\Psi_O\rangle = S|\Psi_I\rangle$, which is governed by the unitary scattering operator $S$. The solid and dashed circles denote the isolated and open system boundaries, respectively. Disks represent optical materials inside the systems. Red and blue colors denote optical fields.

In the following sections, we introduce various analytical and numerical strategies to realize high-dimensional ($N > 2$), universal, and reconfigurable $U_N$ matrices for light waves. The methods are classified by two categories: the systematic assemblies of (i) lower-dimensional unitary matrices, for example, $U_2$ (Section 3), or (ii) the optimized $U_N$ matrices (Section 4). Pros and cons of each approach will be discussed in terms of practical realizations, deterministic designs, programmability, and universality.

## 3 $U_N$ Operations by Assembling Lower-dimensional Units

Given the relation $U_N^\dagger U_N = U_N U_N^\dagger = I$, a $U_N$ matrix has $N^2$ degrees of freedom considering its symmetry and complex-valued components. To handle this large number of variables, a conventional approach has employed the systematic assembly of lower-dimensional unitary units, achieving deterministic design of an arbitrary $U_N$ matrix [61]. The most widely used building block is a $U_2$ matrix in the SU(2) group, which is analogous to the evolution of a single qubit. In this section, we review the basic design strategy (Section 3.1) and its photonic implementations in the spatial (Section 3.2) and temporal (Section 3.3) domains.

### 3.1 Design strategy

In describing light behaviors in both isolated and open systems, a unitary matrix illustrates the evolutions of the couplings between optical elements, as shown in Eqs. (1,2). In this context, a major challenge in constructing higher-dimensional $U_N$ matrices arises from their off-diagonal components, which necessitate the couplings between far-off elements. Because nearby interactions are dominant especially in the platforms for photonic integrated circuits whether exploiting the couplings through propagating [62] or evanescent [45] light, the basic strategy of realizing $U_N$ is to develop the systematic assembly of nearby interactions to effectively achieve far-off couplings. Mathematically, this strategy is based on the conservation of the unitarity under the multiplications, as $(VUW)^\dagger(VUW) = I$, where $U$, $V$, and $W$ are unitary matrices. Therefore, a common objective in this method is to achieve the diagonalization through the multiplications, as $VU_NW = D$, where $D$ is the diagonal matrix that can be obtained simply with isolated optical elements, and $V$ and $W$ represent unitary matrices achievable with nearby couplings. While the target unitary matrix can be reconstructed with the multiplication form $U_N = V^\dagger DW^\dagger$, the resulting matrix multiplication corresponds to the cascaded operations exerted on light, which are executed with a series of varying optical systems along the spatial or temporal axis.

As an example, we introduce the method of implementing an arbitrary $U_N$ using a set of two-dimensional (2D) unitary matrices $U_2$, which was firstly developed in [61]. For simplicity, we restrict our discussion to $U_2$ that reflects only the nearest-neighbor (NN) couplings in one-dimensional (1D) systems, which cover conventional programmable photonic integrated circuits using multiple waveguides [25]. To apply the unitary multiplication, we develop the $N$-dimensional unitary matrix $T_m$ that includes a nontrivial 2D unitary operation $U_2$ at the components $(m,m)$, $(m,m+1)$, $(m+1,m)$, and $(m+1,m+1)$, while the rest components of $T_m$ follow the $N$-dimensional identity matrix $I_N$. The role of $T_m$ is to cancel out an off-diagonal component through the multiplications of $T_mU$ and $UT_m$, which allow for nulling one of the $m$th or $(m+1)$th row and the $m$th or $(m+1)$th column components, respectively (Fig. 2a).

To achieve the diagonalization with a set of $T_m$, $N(N-1)/2$ nulling processes applied to either the upper or lower triangular off-diagonal components are necessary (Fig. 2b,c), because a triangular unitary matrix is a diagonal matrix. Therefore, in the unitary matrix reconstruction $U_N = V^\dagger DW^\dagger$, $V$ and $W$ are constructed with a series of NN-coupling unitaries $T_m$. We note that the sequence of the nulling processes supports design freedom as long as a series of the cascaded nulling processes operate independently, that is, the protection of the nulled components during the subsequent processes. Representative examples include the Reck ([61], Fig. 2b) and Clements ([29], Fig. 2c) designs, which result in the optical hardware with different feature sizes and symmetry conditions (Fig. 2d,e). Notably, owing to its structural symmetry, the Clements design achieves almost half the circuit depth and higher fidelity with error-tolerant interferences, compared to the Reck design.

The range of the unitary group accessible through nulling processes is determined by the universality of the unit unitary matrix $T_m$ and the diagonal matrix $D$. Particularly, with the universal form of $T_m$ and $D$, a universal set of $U_N$ that covers random Haar matrices [63] can be achieved deterministically. Because the diagonal components of $D$ have the unit modulus, the tailored phase evolutions in individual optical elements, such as optical waveguides or resonators with phase shifters, enables the universal form of $D$. Therefore, a major challenge in achieving programmable and universal $U_N$ construction is the realization of the reconfigurable building block for $T_m$, which has to satisfy the com-plete coverage of the SU(2) group, at least, up to the global phase for nulling an arbitrary off-diagonal element during the diagonalization. In the following sections, 3.2 and 3.3, we explore the spatial- and temporal-domain realizations of $T_m$, which possess distinct characteristics in their design procedures and the performance figures of the resulting $U_N$.

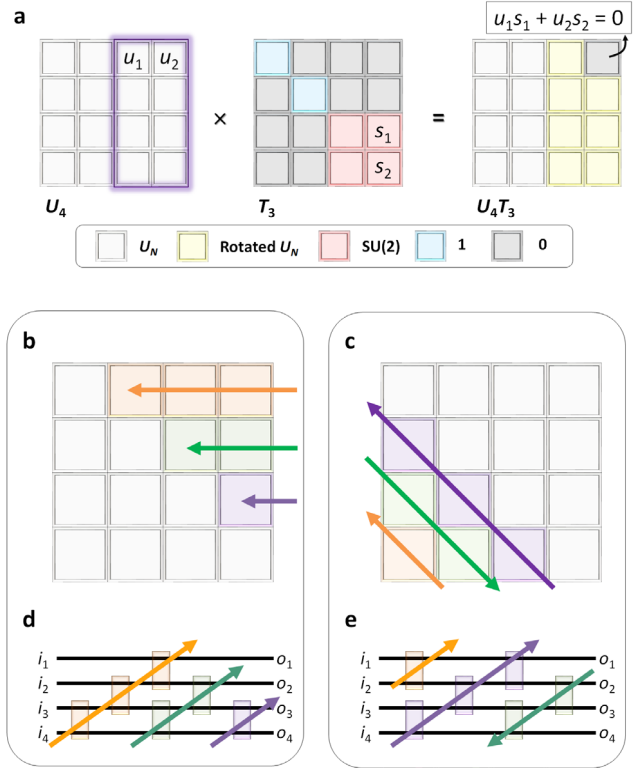

**Figure 2. Universal unitaries composed of lower-dimensional units through nulling processes.** (a) Example of nulling processes in the case of $U_4$ (components with white squares). By multiplying $T_3$ (red squares with SU(2) operations, blue squares with 1, and black squares with 0) on the right side of $U_4$, one of the $U_4 T_3$ components becomes zero (black square) by designing the SU(2) operation to satisfy $u_1 s_1 + u_2 s_2 = 0$. While some of the components in $U_4 T_3$ are identical to those in $U_4$, the other components are the SU(2)-transformed $U_4$ components. The purple box in $U_4$ denotes the SU(2)-transformed part. (b,c) The nulling processes of the (b) Reck [61] and (c) Clement [29] designs, and (d,e) their hardware implementations, respectively. The color of each arrow in (b,c) corresponds to the optical element of the same color in the circuits of (d,e). In (b-e), the nulling process is achieved sequentially, following the order of the orange, green, and purple arrows. The colored boxes in (d,e) represent the SU(2) building block described in Sections 3.2 and 3.3. In (d,e), $i_n$ and $o_n$ are the $n$th input and output signals, respectively ($1 \leq n \leq 4$).

## 3.2 Spatial domain implementation

Because $T_m$ describes the evolution of photonic states in a two-level system up to the global phase, the corresponding operation is characterized by two degrees of freedom: the relative differences in the amplitudes and phases of the fields inside two elements [25]. As discussed in quantum computation [3], the universality of such an operation is guaranteed by a set of two arbitrary rotation operations among $R_x$, $R_y$, and $R_z$, which denote the rotations about $x$-, $y$-, and $z$-axis, respectively. These rotation operations, generally describing the $\theta$-angle rotation about the vector $\mathbf{n}$, can be expressed as the following propagator form:

$$R_n(\theta) = \exp\left(-i\frac{\theta}{2}\mathbf{n}\cdot\boldsymbol{\sigma}\right) \tag{3}$$

where $\boldsymbol{\sigma}$ is the Pauli vector $\boldsymbol{\sigma} = \mathbf{x}\sigma_x + \mathbf{y}\sigma_y + \mathbf{z}\sigma_z$ for the following Pauli matrices:

$$\sigma_x = \begin{bmatrix} 0 & 1 \\ 1 & 0 \end{bmatrix}, \quad \sigma_y = \begin{bmatrix} 0 & -i \\ i & 0 \end{bmatrix}, \quad \sigma_z = \begin{bmatrix} 1 & 0 \\ 0 & -1 \end{bmatrix}. \tag{4}$$

Regarding passive and active optical elements widely used in integrated photonics, the rotation operations that are practically more accessible are $R_x$ and $R_z$. First, as shown in the form of $\sigma_x$ in Eq. (4), which is the generator of $R_x$, the rotation $R_x$ can be obtained with the in-phase coupling between two elements, such as two coupled waveguides composing a directional coupler. The amount of the $R_x$ rotation is then determined by the magnitude of the coupling coefficient and the length of the coupler. Second, the rotation $R_z$ is achieved with $\sigma_z$ in Eq. (4), which is achieved with the decoupled optical elements with different on-site energies such as the propagation constants in waveguides and the resonance frequencies in resonators. To achieve arbitrary rotations of $R_x$ and $R_z$, the listed optical parameters—coupling coefficient, coupler length, propagation constants, and resonance frequencies—need to be reconfigurable. It is worth mentioning that the realization of $R_y$, which requires the gauge field in the coupling due to the imaginary off-diagonal terms in $\sigma_y$, is not straightforward in the spatial-domain integrated photonic platforms. Instead, $R_y$ can be obtained with a series of $R_x$ and $R_z$, as $R_y(\theta) = R_x(3\pi/2)R_z(\theta)R_x(\pi/2)$.

The SU(2) gate, which conducts an arbitrary $U_2 \in$ SU(2) matrix operation for the nulling processes in Fig. 2, can be realized by cascading the constituent operations $R_x$ and $R_z$, as $U_2 = \Pi_k R_{n(k)}(\theta_k)$, where $n(k)$ denotes the rotation axis $x$ or $z$, and $k$ is the order index of the SU(2) operation. It is noted that the configuration of $U_2$, which is determined by the number and order of the rotations, is not unique. Although the simplest form of $U_2$ would be the straightforward multiplication of two orthogonal rotations, such as $U_2 = R_z(\theta_2)R_x(\theta_1)$ or $R_z(\theta_2)R_y(\theta_1)$, one needs to consider the practical implementation of each configuration in terms of the operation fidelity, the method for reconfigurability, and the compatibility with fabrication techniques.

Regarding the difficulty in directly implementing $R_y$, the most straightforward architecture employs tunable directional coupling [64-66] (Fig. 3a,b). The tunable directional coupler corresponds to a reconfigurable $R_x$ rotation, which enables the following relation:

$$U_2 \equiv R_z(\eta)R_x(\gamma) = \begin{bmatrix} e^{-i\frac{\eta}{2}}\cos\left(\frac{\gamma}{2}\right) & -ie^{-i\frac{\eta}{2}}\sin\left(\frac{\gamma}{2}\right) \\ -ie^{i\frac{\eta}{2}}\sin\left(\frac{\gamma}{2}\right) & e^{i\frac{\eta}{2}}\cos\left(\frac{\gamma}{2}\right) \end{bmatrix}, \qquad (5)$$

where $\gamma$ is the rotation angle of $0 \leq \gamma \leq \pi$ about the $x$-axis. Although stable manipulation of the coupling is a critical engineering issue, as discussed in below, the mathematically straightforward structure of the unit and the resulting geometrical simplicity are clear advantages of this approach.

On the other hand, the most widely used configuration involves a fixed $R_x$ (Fig. 3c,d), highlighting the technical challenges in achieving reconfigurability of coupling coefficients. To adjust the coupling for controlling $R_x(\theta)$ without affecting $R_z$ rotations, the optical modes within the elements must be perturbed symmetrically. Furthermore, the nonlinear relationships between material perturbations and modal profiles [67-69], as well as between the perturbed modal profiles and coupling coefficient [70], need to be mitigated. Therefore, to avoid the sensitivity and nonlinearity in performing reconfigurable $R_x(\theta)$, most conventional programmable photonic circuits (PPCs) have leveraged the relationship of $R_x(\pi)R_y(\xi) = R_x(\pi/2)R_z(\xi)R_x(\pi/2)$, which leads to

$$U_2 \equiv R_z(\eta)R_x(\pi/2)R_z(\xi)R_x(\pi/2) = \begin{bmatrix} e^{-i\frac{\eta}{2}}\sin\left(\frac{\xi}{2}\right) & e^{-i\frac{\eta}{2}}\cos\left(\frac{\xi}{2}\right) \\ e^{i\frac{\eta}{2}}\cos\left(\frac{\xi}{2}\right) & -e^{i\frac{\eta}{2}}\sin\left(\frac{\xi}{2}\right) \end{bmatrix} = R_z(\eta)R_x(\pi)R_y(\xi). \qquad (6)$$

As shown in Fig. 3a,b, the corresponding SU(2) gate is composed of two identical passive Mach-Zehnder interferometers both for $R_x(\pi/2)$, and two phase shifters of $\xi$ and $\eta$ for $R_z(\xi)$ and $R_z(\eta)$, respectively. When $0 \leq \xi \leq \pi$ and $0 \leq \eta < 2\pi$, the degrees of freedom provided by two rotations $R_z(\eta)$ and the effective $R_y(\xi)$ allow for the realization of universal SU(2) operations, completely covering the entire evolutions on the Bloch sphere. It is worth mentioning that although other configurations, such as $U_2 = R_z(\eta)R_x(3\pi/2)R_z(\xi)R_x(\pi/2) = R_z(\eta)R_y(\xi)$ that is a mathematically more concise form, are also allowed, the critical advantage of the configuration in Eq. (5) lies in the identical form of $R_x(\pi/2)$, which results in the same fabrication conditions for the Mach-Zehnder interferometers.

To achieve the necessary modulation for the reconfigurable operations of the SU(2) gate, various modulation schemes have been examined. A conventional approach involves using thermo-optic modulation [10, 65, 71-74], which possess advantages of compact gate design due to substantial changes in the refractive index, and low insertion loss. However, this approach presents several challenges, such as limited operation bandwidth of a few MHz and high-power consumption. To address each issue, various alternative solutions have been successfully demonstrated. For example, electro-optical modulation using lithium niobite films offers from 1 to 100 GHz modulation for phase shifts [75, 76]. In terms of power consumption, piezo-optomechanical techniques enable μW-scale operations per modulator with operation speeds about 100 MHz [77, 78]. Owing to the sharply different features of each modulation technique, the method of modulation is a core factor in determining the performance and application target of the entire PPC (Fig. 3e).

All modulation techniques and the passive elements used in the SU(2) gate inevitably involve noise and defects. Furthermore, such random errors scale with $O(N^{1/2})$ due to the $O(N)$ circuit depth. Therefore, addressing fidelity issues through error correction is critical, especially when conducting computing functionalities, even for stochastic operation such as deep learning. To address imperfections in linear-optic elements, such as beam splitters, various systematic calibration procedures have been developed at both the elemental [28] and matrix [79-84] levels, demonstrating that high-fidelity universal unitary operations can be achieved by using additional phase shifts through auxiliary elements or through the tunability of existing ones. While *in situ* training of phase shifts has been demonstrated in deep learning applications [85], recent efforts have successfully reduced the need for internal detection, accurate pre-characterization, and auxiliary elements through unit cell and architecture design [84, 86-88], achieving superior accuracy in tasks, such as pattern recognition (Fig. 3f). In a similar context, a method inspired by network science has recently been proposed, focusing on categorizing active optical elements based on their impacts on achieving reconfigurable universal unitary operations [89]. The work demonstrated that the building blocks of PPCs designed under the Clements approach [29] for achieving random Haar matrices follow the power law and the Pareto principle, stating that most of the significant rotation operations come from about 20% of the building blocks—the 'tail' elements of the circuit. This analysis inspires the pruning of less important phase shifters in the overall circuit (Fig. 3g), which enables the removal of tunable phase shifters that are usually the most expensive elements in PPCs in terms of noise and footprint engineering. As shown in Fig. 3h, pruning the 'body' of the phase shifters—less important ones, constituting about 80% of the circuit—can yield even better performance at a specific level of noise, aligning with the underlying mechanisms of hub nodes in network science [90]. Recently, the concept of pruning for photonic circuits has also been employed in relation to the training process [91, 92], achieving reduced power consumption and noise resilience.

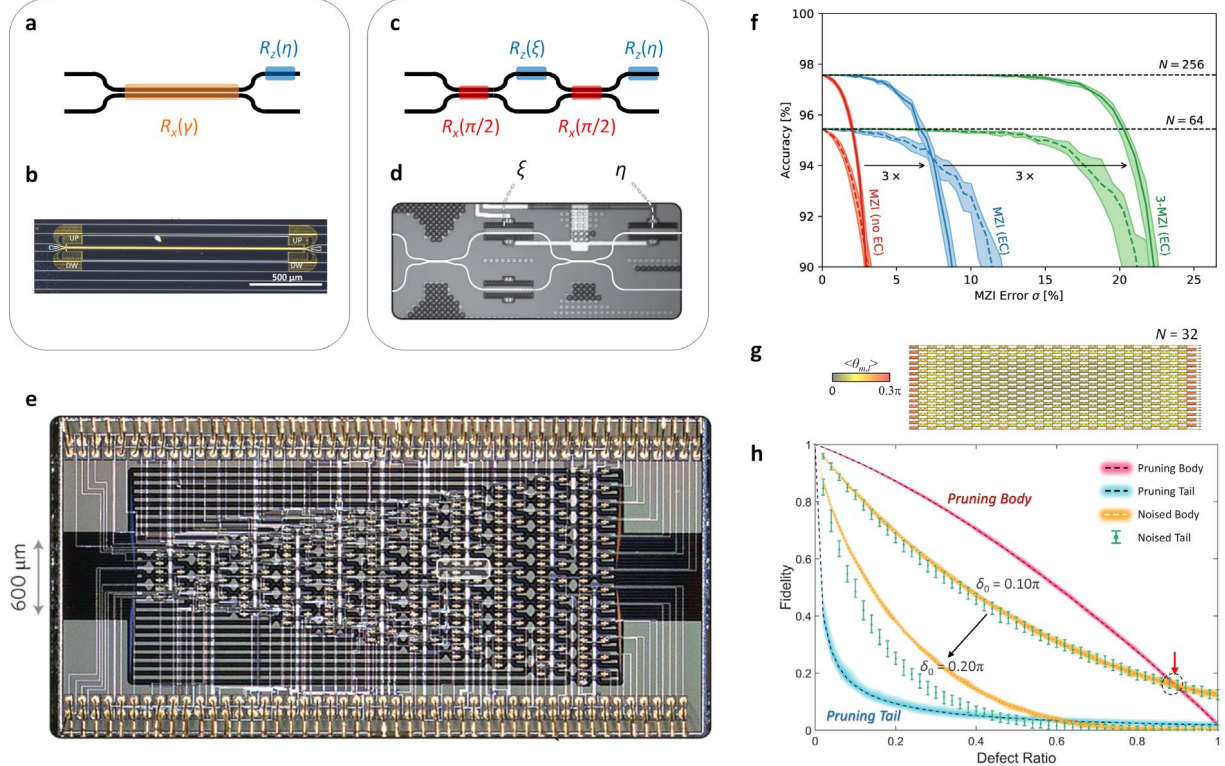

**Figure 3. Spatial implementation of universal unitary circuits.** (a,b) SU(2) gate with the reconfigurable $R_x$ rotation: (a) schematic of the gate, and (b) its real implementation [64]. (c,d) SU(2) gate with the fixed $R_x$ rotations: (c) schematic of the gate, and (d) its real implementation [10]. (e) The PPC implementation using a set of SU(2) gates in (d). (f) Effect of hardware errors on pattern recognition [88] for 28 × 28 images from the MNIST dataset [93]. 'EC' denotes the error correction achieved by the self-configuration of Mach-Zehnder interferometers. (g,h) The pruning of PPCs for noise-resilient unitary operations [89]. (g) The averages of the phase shift $\xi = \langle\theta_{m,l}\rangle$ for the PPCs of 100 $U_{32}$ realizations. (h) Comparison of the fidelities of the $U_{128}$ PPCs in different groups: pruning body (red line), pruning tail

(blue line), noisy body (orange line), and noisy tail (green error bars). The thicknesses of the colored lines and the error bars present the range of the fidelities between their maxima and minima. Panel (b) reprinted with permission from [64], Optica. Panels (d,e) reprinted with permission from [10], Springer Nature Limited. Panels (f) and (g,h) are adapted with permission from [88] and [89], respectively, Springer Nature Limited.

## 3.3 Temporal domain implementation

Despite the great success of spatial-domain PPCs especially in terms of fidelity [28, 31, 79-89, 94], scalability still remains another critical challenge. To diagonalize a $U_N$ matrix using SU(2) operations, $N(N-1)/2$ nulling processes are necessary as described in Fig. 2. Even when approximately $N/2$ operations are achieved in parallel, as is available in the Clements design [29], the circuit depth is proportional to $N$, which results in $O(N^2)$ scalability in both device footprint and gate number regarding the channel number $N$. This poor scalability, for example, requiring over $10^6$ elements with footprints exceeding a few cm$^2$ for a deep learning model with only 1,000 neurons, also raises substantial fidelity issue due to imperfections in device fabrication and light sources.

One possible approach to overcoming this limitation is to exploit the mathematical similarity between the governing equations in different physical axes. For example, the spatial propagation of light along a waveguide and the temporal evolution inside a resonator can be described by the almost same equations of $da/dx = -i\beta a$ and $da/dt = +i\omega a$ [70], respectively, where $a$ is a field amplitude, $\beta$ is the wavenumber along the waveguide, and $\omega$ is the resonance frequency of the resonator. Therefore, one can replace the field evolution of light along the spatial domain with the evolution along an alternative physical axis—time or frequency—which allows for the substitution of the device footprint with the operation time or mode number. Similar approaches have been employed in examining synthetic-dimensional phenomena [13, 95-100], supersymmetric transformation [101], crystal optics [39, 102], and disordered photonics [41, 43].

The proposal of programmable photonic time circuits (PPTCs) [103] corresponds to the space-to-time replacement in the realization of reconfigurable photonic circuits, aiming to address the poor footprint scalability of PPCs by utilizing time-domain computation. In PPTCs, the nulling process for diagonalization is identical to the spatial counterpart, while the spatial SU(2) gate is replaced with the temporal one. This SU(2) 'time gate' consists of two identical travelling-wave resonators with resonance frequency $\omega_0$ (Fig. 4a) to describe two-level systems with the spinor state $\Psi = [\psi_m, \psi_n]^T$, where $\psi_m$ and $\psi_n$ are the field amplitudes of the specific pseudospin modes in the neighboring $m$th and $n$th resonators, respectively. The resonators are coupled via a pair of single-mode waveguide loops with a decay rate $1/\tau$, which have been widely employed in topological photonics to achieve pseudogauge fields in coupling [34, 104, 105]. When preparing tunable phase shifters in the resonators and waveguide loops, the spinor state is governed by the following spinor Hamiltonian [103]:

$$H_S = -\omega_0 \sigma_0 - \frac{1}{2\tau}\left[\cos\xi^U(t) + \cos\xi^L(t)\right]\sigma_x - \frac{1}{2\tau}\left[\sin\xi^U(t) + \sin\xi^L(t)\right]\sigma_y - \Delta\omega(t)\sigma_z, \qquad (7)$$

where $\xi_{mn}^U(t)$ and $\xi_{mn}^L(t)$ are the time-varying phase shifts in the upper and lower loops, and $\Delta\omega$ is the time-varying averaged resonance perturbation due to the tunable phase shifters in each resonator. Equation (7) shows that the full degrees of freedom for the rotation operations, $R_{x,y,z}$, are accessible with the SU(2) time gate. By imposing the digital modulation on tunable phase shifters, the rotation operations about an arbitrary axis on the $xy$-plane (Fig. 4b) and about the $z$-axis (Fig. 4c) can be realized, comprising a universal set for the unitary operation of a spinor state. Such a universal SU(2) operation facilitates the extension to universal and reconfigurable $U_N$ operations following the method described in Section 3.1, as demonstrated in the realization of the quantum Fourier transform and random Haar matrices [103], and the emulation of light behaviors inside hyperbolic lattices [106]. Because the computation is achieved with light stored inside coupled resonators, the PPTC can be understood as compressing the spatial circuit depth from $N$ to 1, by transferring the domain for necessary computations to the temporal axis. The design strategy with tunable gauge fields has recently been extended to the implementation of integrated photonic platforms for reconfigurable matrix-valued gauge fields and the following programmable non-Abelian physics and braiding [107], which revealed novel topological phases in photonic lattices.

Remaining issues in the practical implementation of PPTCs lie in the design of modulation signals and the limitations in modulation speed. Because Eq. (7) is nonlinear, the calculation of the required phase shifts generally does not yield an analytical solution. In [103], digital modulation is applied as an analogy of the discretized building blocks in spatial-domain PPCs. However, because the speed of temporal modulation is limited by the response time of materials, ideal discretization in the PPTC is actually unattainable, and therefore, the fidelity significantly degrades with slower modulation of phase shifts (Fig. 4d). The solution to this issue can be found in digital signal processing techniques [103], such as predistortion [108], and analog computation [109-113].

A more critical issue would be the necessity of ultrahigh quality factors that should guarantee the storage of light during modulation. Notably, because the SU(2) gate is executed sequentially in previous works [103, 106] due to the lack of synchronization between SU(2) time gates, the resulting scalability of the necessary time period scales as $O(N^2)$. The realization of parallel SU(2) operations, analogous to the spatial implementation [29, 61], can substantially mitigate this problem, facilitating $O(N)$ scalability. Dynamical control of coupling quality factors [114] can also provide an optimum solution by restricting unwanted dissipation to the intrinsic loss of resonators.

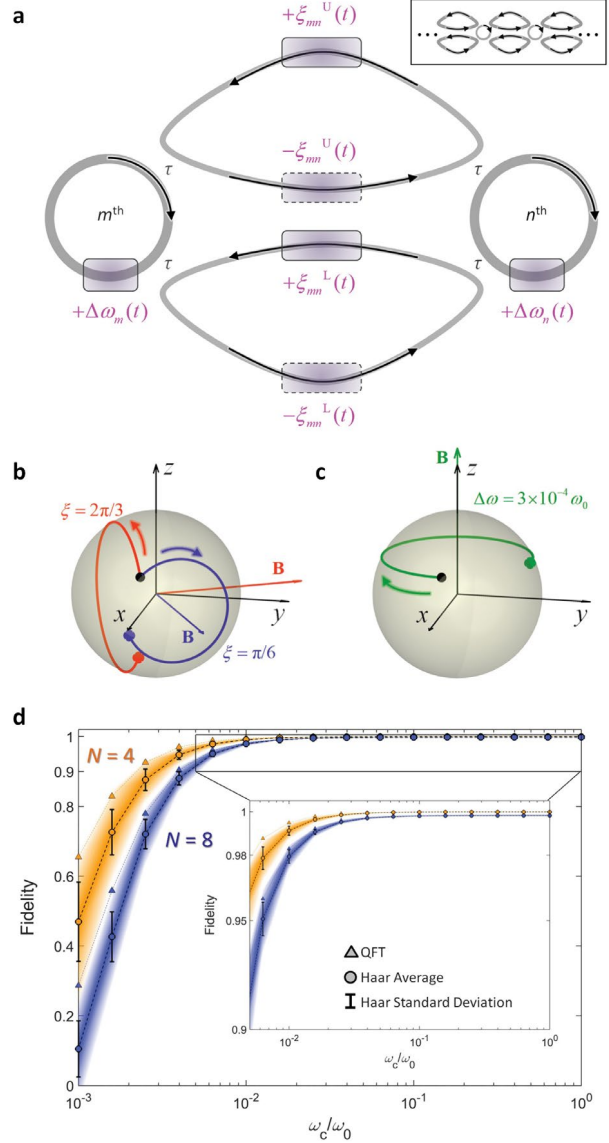

**Figure 4. Temporal implementation of universal unitary circuits.** (a) The SU(2) time gate between the $m$th and $n$th channels. Gray circles and curved triangles represent resonators and waveguide loops, respectively. Purple boxes are phase shifters. The upper right inset illustrates a PPTC comprised of multiple SU(2) time gates. (b,c) Temporal evolution of pseudospinor modes on the Bloch sphere for two operation modes: (b) even-parity mode for $\xi_{mn}^U(t) = \xi_{mn}^L(t) = \xi$ and $\Delta\omega(t) = 0$ in (a), and (c) odd-parity mode for $\xi_{mn}^U(t) = -\xi_{mn}^L(t) = \pi/2$ and $\Delta\omega(t) \neq 0$ in (a). **B** denotes the pseudospinor rotation axis. (d) PPTC fidelities with respect to the cutoff frequency $\omega_c$ of low-pass filtering. Circles and error bars represent the average and standard deviation of 20 realizations of random Haar matrices at each $\omega_c$ and triangles represent quantum Fourier transforms for $N = 4$ (two qubits; orange) and $N = 8$ (three qubits; blue). Panels (a-d) reprinted with permission from [103], APS.

# 4  $U_N$ Operations by Cascading Higher-Dimensional Cells

Although the assembly of lower-dimensional unitary units allows for the deterministic implementation of a higher-dimensional $U_N$ matrix in photonic systems, the resulting architecture of the circuit is nonunique, even when we employ an identical form of unitary gates, as demonstrated in various previous works [29, 30, 61]. Therefore, the assembly introduced in Section 3 does not guarantee the optimal design for practical implementation in terms of the device footprint, gate number, signal fidelity, and the geometrical complexity of the design. This limitation inspires the necessity of developing optimization techniques to explore a global optimum by simultaneously accessing the full $N^2$ degrees of freedom $U_N$. In this section, we review a representative example in this field—the realization of high-dimensional universal unitary operations through the cascaded unit cells of the same dimension—focusing on the basic design strategy (Section 4.1), and the implementations in the spatial (Section 4.2) and frequency (Section 4.3) domains.

## 4.1  Design strategy

Although the detailed design process described in this section is distinct from that in Section 3, the underlying philosophy is identical: constructing the entire $U(N)$ group with $N^2$ degrees of freedom from the layers of $U(N)$ subsets that are physically allowed in terms of nearby interactions. The difference lies in the dimensionality of the nontrivial parts of $U(N)$ layers. In Section 3, we introduced the use of two-dimensional SU(2) operations while the other channels experience U(1) evolution. In this section, we focused on the use of $N$-dimensional operations, which can be implemented through nearby interactions. Consider the realization of $N$-dimensional universal unitary operations $U$ using the $N$-dimensional cascaded unitary unit cells $V_m$ ($m = 1, 2, \ldots, M$; Fig. 5a), as $U = V_M V_{M-1} \cdots V_2 V_1$, while $V_m$ does not need to be universal. The first step in covering $N^2$ degrees of freedom of $U$ is to devise a photonic unit cell that provides $V_m$ through nearby interactions, because achieving long-range interactions is challenging on conventional integrated platforms. By performing cascaded multiplications of $V_m$, we can then obtain the corresponding photonic hardware that approximates $U$ through the substantially enhanced degrees of freedom.

In this approach toward universal unitaries, the first example of $V_m$ is the unit cell that conducts Fourier-transform operations in the spatial domain. Various photonic systems can operate as gates for the Fourier transform of optical information, which is a subclass of $N$-dimensional unitary operations deterministically accessible with nearby interactions between optical elements. A representative example is a diffractive optical system, where light propagation in the Fraunhofer regime leads to the Fourier transform of the transverse profile of an optical field [38]. As a theoretical background, it was demonstrated that a reconfigurable $y$-axis rotation operation on the Bloch sphere of an arbitrary pair of channels can be achieved by using an $N$-channel discrete Fourier transform in conjunction with an $N$-channel diagonal unitary operation [57]. Because the diagonal unitary operation permits tunable $z$-axis rotations for any pair of channels, the combination of a set of Fourier-transforming ($U_{FT}$) and diagonal ($D$) unitary operations enables the configuration of the universal SU(2) operation for each pair of channels, as well as the subsequent $N$-dimensional unitary operations as demonstrated in Section 3 (Fig. 5b).

It is important to note that the Fourier transform is not the only high-dimensional configuration suitable for constructing universal unitaries. Any form of reflectionless transfer matrix can serve as the unit cell for composing universal $U_N$ (Fig. 5c). Although such transfer matrices in integrated platforms are typically characterized by nearby interactions, long-range interactions can also be achievable using specific platforms. A particularly interesting example involves employing the concept of synthetic dimensions [13, 95-100] to construct the unit transfer matrix using higher-order couplings. This approach also replaces spatial degrees of freedom with the frequency axis for enhanced scalability in footprint and element number, as similar to the approach in Section 3.3.

In the following sections, 4.2 and 4.3, we explore the realizations of $V_m$ using the Fourier transform in the spatial domain and the transfer matrix in the frequency synthetic-dimensional domain, discussing their theoretical background, numerical assessment, and practical implementation.

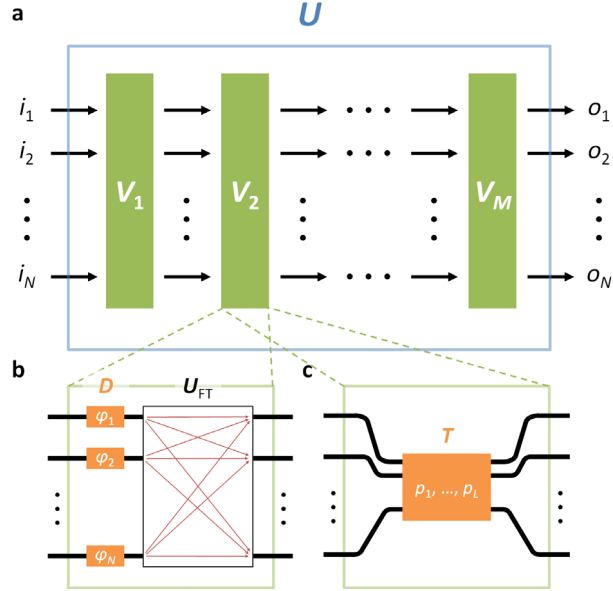

**Figure 5. Universal unitaries composed of higher-dimensional cells through numerical optimization.** (a) The circuit realizing universal unitary operations $U$. $i_n$ and $o_n$ are the $n$th input and output signals, respectively ($1 \leq n \leq N$). $V_m$ is the reconfigurable $N$-dimensional unitary unit cell ($1 \leq m \leq M$) of which the operation is determined by nearby interactions accessible with integrated photonic platforms. (b,c) Detailed implementation of $V_m$ cell. (b) $V_m$ composed of the diagonal unitary operation $D$ and the Fourier transform $U_{FT}$. The diagonal unitary matrix is realized with an array of phase shifters: $\varphi_n$ for the phase shift applied to the $n$th channel. (c) $V_m$ realized through the transfer matrix $T$, which depends on $L$ real-valued parameters $p_l$ ($1 \leq l \leq L$).

## 4.2 Spatial Fourier implementation

In integrated photonics, which leverages nearby interactions between guided lights in 2D platforms, the Fourier transform is implemented upon the concept of the fractional Fourier transform (FrFT) [115], a generalization of the standard Fourier transform. Because the Fourier transform operator and its generalization, the FrFT operator, are both unitary, their physical implementations rely on exploring Hermitian Hamiltonians of which the propagators correspond to these operators. Consider the following operator representation of the 1D Fourier transform:

$$F[f(x)] = \frac{1}{\sqrt{2\pi}} \int_{-\infty}^{\infty} f(x') e^{-ixx'} dx', \tag{8}$$

where the conventional expression of the Fourier transform of $f(x)$ is obtained by applying $F$ to $f(x)$, and then, replacing $x$ at the right side of the equation with the wavenumber $k$. The underlying concept of the FrFT involves interpreting the Fourier transform operator $F$ as $F = F(\pi/4)$, as described in below. As shown in [115], the operator $F$ satisfies the following eigenvalue equation:

$$F e^{-x^2/2} H_n(x) = e^{-in(\pi/2)} e^{-x^2/2} H_n(x), \tag{9}$$

where $H_n(x)$ is the Hermite polynomial of order $n$. Notably, the eigenmodes in Eq. (9) are familiar in quantum mechanics, because they are the eigenmodes of the harmonic oscillator Hamiltonian:

$$H = -\frac{d^2}{dx^2} + x^2 - 1, \tag{10}$$

while the eigenvalue of the Hamiltonian is $2n$. Regarding that $F$ is a unitary operator, the Fourier transform can be considered the unitary propagator of the harmonic oscillator Hamiltonian according to the Stone's theorem [2]. From Eq. (1), the propagator applied to the $n$th eigenmode is $\exp(-iH\xi) = \exp(-2in\xi)$, which leads to $\exp(-in\pi/2)$ at $\xi = \pi/4$. Because this relationship is established for all the eigenmodes independent of $n$, the Hamiltonian of Eq. (10) provides the Fourier transform of an arbitrary state $\psi(\xi_0)$ exactly at $\psi(\xi_0 + \pi/4)$ from Eq. (9), allowing for introducing the relation $F = F(\pi/4)$. A generalized propagator $F(\xi)$, which leads to $\exp(-iH\xi)$ for the Hamiltonian $H$ in Eq. (10), then satisfies:

$$F(\xi)e^{-x^2/2}H_n(x) = e^{-i2n\xi}e^{-x^2/2}H_n(x). \quad (11)$$

This unitary propagator $F(\xi)$ corresponds to the FrFT operator, which includes the Fourier transform as the special case when $F(\xi = \pi/4)$. The graphical illustration of Eq. (11) in phase space is shown in Fig. 6a.

Equations (10) and (11) demonstrate that the optical implementation of a harmonic oscillator potential enables light-based computing of both the Fourier transform and FrFT. According to the mathematical similarity between the Schrödinger equation and the optical paraxial wave equation [6, 7], the harmonic oscillator potential can be readily obtained by engineering an effective refractive index profile [116-118]. Because a higher refractive index corresponds to a lower quantum potential, the index profile of optical harmonic oscillators is shaped like a convex lens, with a higher index around the center.

In integrated photonics, achieving a continuous variation of the refractive index could be challenging because the unit elements, such as optical waveguides or resonators, are usually discretized with fixed geometrical and material specifications. Therefore, a common approach lies in realizing the discrete fractional Fourier transform (DFrFT) [119], which corresponds to the generalization of a discrete Fourier transform. Because harmonic oscillators can also be effectively reproduced in discretized systems, such as waveguide arrays with detuned coupling or propagation vectors [120-122], the DFrFT and its special case of the discrete Fourier transform can be implemented by tailoring the evolution parameter $\xi$ within optical harmonic oscillators (Fig. 6b) [123].

The DFrFT on integrated photonic platforms has been intensively studied to realize light-based computing by performing universal unitary operations [124, 125], convolution operations [126], and arbitrary linear operations [127] (Fig. 6c-e). Although it has been demonstrated that universal unitary operations can be obtained through alternating multiplications of diagonal unitary matrices and Fourier-transforming ($U_{FT}$) matrices [57], designing the diagonal matrices and determining the number of unit cells (*M* in Fig. 5a and Fig. 6d,e) still requires numerical assessments. Recently, a more general form, involving the DFrFT instead of the discrete Fourier transform, was used to obtain universal unitary operations [124]. In this approach, numerical optimization with the Levenberg-Marquardt algorithm [128, 129] was employed to inversely design diagonal unitary matrices, which correspond to the phase-shifting layers (*D* in Fig. 5b). The result demonstrates that (*N*+1) phase-shifting layers are sufficient to reproduce universal unitary operations approximately, resulting in $O(N^2)$ scalability both for the footprint and element number, similar to the SU(2)-based design in Section 3.2. Although the design has yet to demonstrate superior scalability, its more simplified geometry without Mach-Zehnder interferometers could be advantageous especially in large-scale realizations. Furthermore, because the design is defect-resilient, maintaining the universality with one faulted phase shifter at each layer, an extended comparison with the SU(2)-based design will be necessary in future studies. It is worth mentioning that alternating multiplications of the Fourier transform, a complex-valued diagonal matrix, and the inverse Fourier transform allow for the construction of circulant matrices [130], enabling the realization of an arbitrary complex matrix [131]. In this context, the approach in [124] has recently been generalized to universal linear operations using the DFrFT [127], which can be directly applied to weight matrices in photonic deep learning. In terms of resolving scalability, the utilization of the FrFT in the time domain [132] for constructing unitary operations can be a topic of future studies, as the correspondence of the relationship between PPCs in Section 3.2 and PPTCs in Section 3.3.

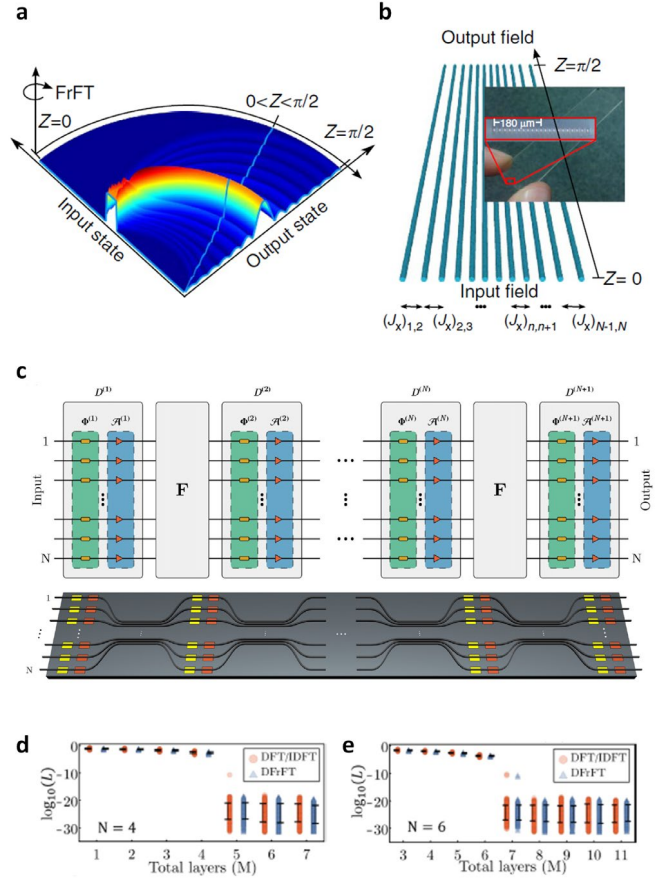

**Figure 6. Spatial Fourier implementation of universal unitary circuits.** (a) Pictorial view of FrFT in phase space, where $Z = 2\xi$ in Eq. (11). (b) A schematic of an engineered waveguide array for an optical harmonic oscillator potential. (c) Schematic for the proposed decomposition of universal linear $N \times N$ matrices with the complex diagonal matrices $D^{(m)}$ and the DFrFT matrix F. The diagonal matrices are further decomposed into a layer of phase shifters $\Phi^{(m)}$ and a layer of amplitude modulators $A^{(m)}$. Schematic of a photonic-device architecture is shown in the bottom figure. (d,e) Errors $L$ in reproducing universal linear matrices for (d) $N = 4$ and (e) $N = 6$, while varying the total number of layers $M$. Panels (a,b) reprinted with permission from [123], Springer Nature Limited. Panels (c-e) reprinted with permission from [127], APS.

## 4.3 Synthetic-dimensional implementation

Despite the simplified and defect-resilient design using the spatial-domain DFrFT, the overall scalability of the footprint and element number still remains $O(N^2)$, which hinders large-scale implementations required for deep-learning applications. Similar to the utilization of the temporal axis in [103], employing computation on alternative physical axes instead of the spatial domain also became necessary in unitary operations using higher-dimensional unit cells. One approach is to utilize the synthetic frequency dimension [97, 133].

The concept of the synthetic dimension allows for exploring light behaviors in higher-dimensional spaces beyond the spatial dimension by exploiting internal degrees of freedom of photons or imposing extended system parameters on the governing Hamiltonian [97]. Figure 7a illustrates an example of the first approach: the use of multimode optical resonances within a single ring resonator, which extends a zero-dimensional (0D) photonic structure to an effective 1D or higher-dimensional system. The key mechanism involves developing coupling between resonance modes, which are initially orthogonal to each other in both the spatial and temporal domains. Such coupling can be achieved through spatio-temporal modulation to break this orthogonality, for example, by finite-space modulation for broken spatial orthogonality [134], and at the same time, by designed temporal modulation to control frequency-domain coupling distributions [95, 135-137]. In this field, electro-optical modulators (EOM) have been widely employed for their sufficiently large modulation speed and depth, which determine the accessible range of the free-spectral range (FSR), the accessible orders of neighboring couplings, and the coupling strength. The synthetic dimension has been widely applied in topological photonics, permitting the observation of three- [13] or even four-dimensional [133] topological phenomena in 2D integrated photonic platforms, with the experimental verification of the concept [98, 138].

Employing the extended dimension from the concept of the frequency-synthetic dimension, $O(N)$-scalable implementation of universal unitary operations has been recently demonstrated (Fig. 7b) [139], which can be generalized to arbitrary linear transformations. In the proposed system, superior scalability is achieved by compressing the spatial circuit width from $N$ to 1, which is obtained by replacing $N$-channel operations with a higher-dimensional $N$-element lattice unit cell developed in the 0D resonator. Each basis vector for matrix operations corresponds to the resonance mode, $\omega_m = \omega_0 + m\Omega_R$, where $m$ is an integer, $\omega_0$ is the reference frequency, and $\Omega_R$ denotes the FSR. To configure the lattice, the EOM modulates the permittivity in the finite region of the resonator through a combination of multiple harmonics, as $\Delta\varepsilon = \Sigma_l \delta\varepsilon_l \cos(l\Omega_R t + \theta_l)$ ($1 \leq l \leq N_f$), where $\delta\varepsilon_l$ and $\theta_l$ are the modulation depth and phase for the modulation frequency $l\Omega_R$, respectively. We note that $l$ values greater than 1 enable long-range couplings, which are one of the critical advantages of utilizing synthetic dimensions compared to the intricate design required for spatial long-range coupling [140]. Such long-range couplings allow for the efficient access to off-diagonal elements in matrix operations, providing higher-dimensional unit cells with excellent expressivity for constructing universal unitaries. The dimensionality $N$ of unitary operations is defined by establishing spectral boundaries that block undesired couplings into too higher- or lower-modes, which can be realized with inner auxiliary rings. The rings split the modes outside the spectral boundaries through the coupling process, which deviates eigenfrequencies, and thus, results in decoupling (Fig. 7c) [98].

To fulfill the necessary degrees of freedom required for unitary operations, the entire system consists of an array of $N_r$ unit cells, each performing an $N$-dimensional unitary transfer matrix $T$ (Figs. 5c and 7b). The unit cell is composed of a synthetic-dimensional ring resonator side-coupled to an external waveguide. Across each unit cell, the waveguide supports the $N\times 1$ input mode vector $s^+ = [\ldots, s_{-1}^+, s_0^+, s_1^+, \ldots]^T$ and the $N\times 1$ output mode vector $s^- = [\ldots, s_{-1}^-, s_0^-, s_1^-, \ldots]^T$, where $s_m^+$ and $s_m^-$ are the input and output modes, respectively, which are coupled with the resonance mode at $\omega_m$. The unit cell then conducts a unitary operation of $s^- = Ts^+$. Notably, the circuit depth $N_r$ is determined by the entire degrees of freedom in multiple harmonic modulations $2N_f$ from $\delta\varepsilon_l$ and $\theta_l$, and the dimensionality $N$, enforcing the condition of $2N_f N_r \geq N^2$ for universality. As similar to the optimization process in Section 4.2, this synthetic-dimensional resonator array is inversely designed using a gradient-based procedure: the limited-memory Broyden-Fletcher-Goldfarb-Shanno algorithm [141]. In contrast to the spatial- or temporal-domain realizations treated in Sections 3.1, 3.2, and 4.2, the unitary transformations in this system are non-reciprocal due to the spatio-temporal modulations [142, 143]. An arbitrary linear transformation was also examined by exploiting the sub-dimensional space of the system for unitary operations [144].

When compared with the PPTC [103], which also achieves $O(N)$ scalability through the transfer of the circuit depth into the temporal axis, the synthetic-dimensional approach that transfers the circuit width to the frequency axis allows for a more mitigated condition on computing time [139], because each resonator needs to store light during $1/N$ part of the entire computation. Nonetheless, we note that the synthetic-dimensional system requires equally-spaced multi-wavelength sources and detectors due to its operation principles, which could compose a challenge in the extension to large-scale integrated systems.

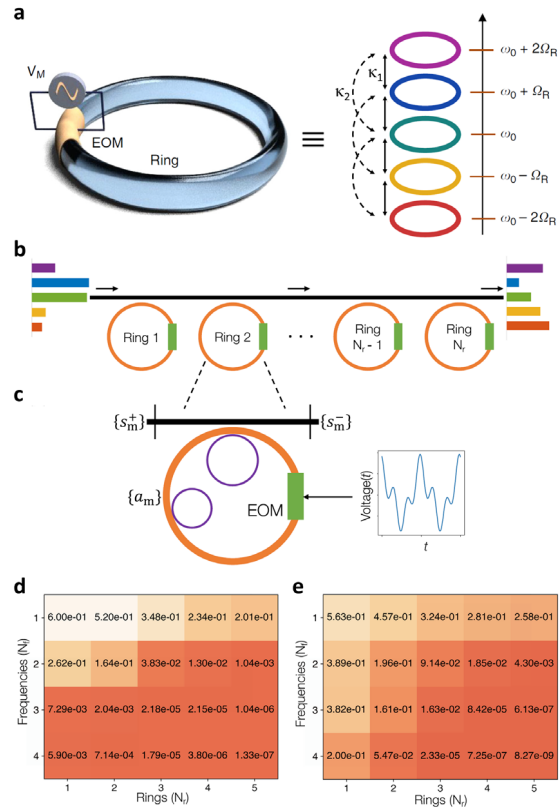

**Figure 7. Synthetic-dimensional implementation of universal unitary circuits.** (a) Dynamically modulated ring resonator and its band structure for comprising the synthetic frequency dimension. The resonator is modulated via the multiple harmonic signal $V_M$ applied to a part of the ring (left), which can create nearest-neighbor ($\kappa_1$) or long-range ($\kappa_2$) coupling (right). 'EOM' denotes an electro-optic phase modulator. (b) The entire system for generating universal unitary operations and the subsequent arbitrary linear transformations. The input is encoded in five-dimensional frequency modes. (c) A unit cell design. The dielectric property of the ring resonator is controlled through the multiple harmonic EOM signal, which allows for the reconfigurable unitary transfer matrix applied to the input profile $\{s_m^+\}$. The purple rings represent the internal auxiliary rings for creating frequency boundaries. (d,e) The errors as functions of the number of the unit cells $N_r$ and the number of modulation harmonics $N_f$ in obtaining a five-dimensional (d) permutation matrix and (e) Vandermonde matrix. Panels (a) reprinted with permission from [138], Springer Nature Limited. Panels (b-e) reprinted with permission from [139], Springer Nature Limited.

## 5 Conclusions and Outlook

After emphasizing the ubiquitous nature of unitary operations in photonics, we explored the design methods for photonic unitary circuits, which enable ultrafast, reconfigurable, and universal unitary operations necessary for light-based computing. We categorized these methods into two groups based on the dimensionality of the building block operations—analytical design using lower-dimensional ones, such as SU(2) operations, and numerical optimization using higher-dimensional ones, such as Fourier transforms. The methods of each group possess different pros and cons.

When lower-dimensional operations, especially SU(2) ones, are employed to compose unitaries, theoretically perfect forms of unitary operations are accomplished systematically. Therefore, this approach allows for more precise computing applications without theoretical defects in principle, enabling the selective engineering of unit cells, such as the calibration to experimental defects [94] and the design of the fault-tolerant geometry [88, 89]. However, because the method itself does not guarantee the optimized configuration in terms of the fidelity, device footprint, and element number, employing a more superior geometry [29, 30] or utilizing other wave degrees of freedom [103] remains an ongoing research topic.

In comparison, the approaches employing higher-dimensional building blocks inevitably utilize numerical recipes due to extended parameter spaces. Therefore, the realization of the perfect fidelity becomes more challenging than the SU(2)-based approach, especially when considering the possible existence of local minima. However, as shown in the DFrFT unit cell [124, 125, 127], this approach can provide more simplified geometry that could be preferred in real implementations. Furthermore, the Fourier-based realization can offer more efficient routes to advanced functionalities, such as convolutional operations [126] and Fourier-space computation [145].

Remaining critical issues in performing unitary-based computations with integrated photonic hardware include achieving scalability below $O(N)$, ensuring high fidelity against defects, and exploring global optima in the circuit design. First, as demonstrated by approaches exploiting temporal [103] and frequency [139] axes, potential solutions for scalability may be found in utilizing the rich internal degrees of freedom of light, such as polarizations and orbital angular momenta. In this context, the application of metamaterials to integrated photonics, such as metatronics [110, 146], can offer a novel solution for increasing the overall information capacity within a limited footprint. Another route may involve the construction of complex network architecture for light [147, 148], which can provide more efficient access to off-diagonal elements through far-off coupling, thereby enabling a compact design. Second, many recent studies have focused on achieving unitary operations with partially coherent light [149-152], which accommodates more practical scenarios. In addition to the use of feedback-based self-configuring systems to manage partial coherence [152], researches on dynamical stable theory [153-156] and topological phenomena [105, 157] in view of unitary operations could be a future research topic to achieve internal fault tolerance as analogous to topological computation by anyonic operations [158]. Finally, in leveraging higher-dimensional unit cells to optimize photonic unitary circuits, the expanded system dimensionality and increased coupling complexity substantially raise the difficulty of finding a global optimum within the parameter space. Although gradient-based approaches, including deep learning for photonic system design [159-163], offer rapid convergence to local minima, designing large-scale photonic circuits may require methods better suited for finding a global optimum, such as neuroevolution [164, 165], particle swarm optimization [166], and a global optimizer using a conditional generative neural network [167].

**Author contribution:** All authors contributed to the final manuscript.

**Acknowledgements:** We thank Beomjoon Chae for valuable discussions. We acknowledge financial support from the National Research Foundation of Korea (NRF) through the Basic Research Laboratory (No. RS-2024-00397664), Innovation Research Center (No. RS-2024-00413957), Young Researcher Program (No. 2021R1C1C1005031), and Midcareer Researcher Program (No. RS-2023-00274348), all funded by the Korean government (MSIT). This work was supported by Creative-Pioneering Researchers Program and the BK21 FOUR program of the Education and Research Program for Future ICT Pioneers in 2024, through Seoul National University. This work was also supported by the 2024 Advanced Facility Fund of the University of Seoul for Xianji Piao. We also acknowledge an administrative support from SOFT foundry institute.

**Conflict of interest statement:** The authors declare no competing financial interest.

20100. C. Qin, S. Wang, B. Wang, X. Hu, C. Liu, Y. Li, L. Zhao, H. Ye, S. Longhi, and P. Lu, "Temporal Goos-Hänchen Shift in Synthetic Discrete-Time Heterolattices," Phys. Rev. Lett. **133**, 083802 (2024).
101. C. García-Meca, A. M. Ortiz, and R. L. Sáez, "Supersymmetry in the time domain and its applications in optics," Nat. Commun. **11**, 813 (2020).
102. X. Wang, M. S. Mirmoosa, V. S. Asadchy, C. Rockstuhl, S. Fan, and S. A. Tretyakov, "Metasurface-based realization of photonic time crystals," Sci. Adv. **9**, eadg7541 (2023).
103. X. Piao, S. Yu, and N. Park, "Programmable photonic time circuits for highly scalable universal unitaries," Phys. Rev. Lett. **132**, 103801 (2024).
104. M. Hafezi, S. Mittal, J. Fan, A. Migdall, and J. M. Taylor, "Imaging topological edge states in silicon photonics," Nat. Photon. **7**, 1001-1005 (2013).
105. T. Ozawa, H. M. Price, A. Amo, N. Goldman, M. Hafezi, L. Lu, M. C. Rechtsman, D. Schuster, J. Simon, and O. Zilberberg, "Topological photonics," Rev. Mod. Phys. **91**, 015006 (2019).
106. H. Park, X. Piao, and S. Yu, "Scalable and Programmable Emulation of Photonic Hyperbolic Lattices," ACS Photon. **11**, 3890-3897 (2024).
107. G. Kim, J. Li, X. Piao, N. Park, and S. Yu, "Programmable lattices for non-Abelian topological photonics and braiding," arXiv preprint arXiv:2410.01181 (2024).
108. J. G. Proakis, *Digital signal processing: principles, algorithms, and applications, 4/E* (Pearson Education India, 2007).
109. F. Zangeneh-Nejad, D. L. Sounas, A. Alù, and R. Fleury, "Analogue computing with metamaterials," Nat. Rev. Mater. **6**, 207-225 (2021).
110. M. Miscuglio, Y. Gui, X. Ma, Z. Ma, S. Sun, T. El Ghazawi, T. Itoh, A. Alù, and V. J. Sorger, "Approximate analog computing with metatronic circuits," Communications Physics **4**, 196 (2021).
111. H. Goh and A. Alù, "Nonlocal scatterer for compact wave-based analog computing," Phys. Rev. Lett. **128**, 073201 (2022).
112. A. Silva, F. Monticone, G. Castaldi, V. Galdi, A. Alù, and N. Engheta, "Performing mathematical operations with metamaterials," Science **343**, 160-163 (2014).
113. S. K. Vadlamani, D. Englund, and R. Hamerly, "Transferable learning on analog hardware," Sci. Adv. **9**, eadh3436 (2023).
114. Y. Tanaka, J. Upham, T. Nagashima, T. Sugiya, T. Asano, and S. Noda, "Dynamic control of the Q factor in a photonic crystal nanocavity," Nat. Mater. **6**, 862-865 (2007).
115. V. Namias, "The fractional order Fourier transform and its application to quantum mechanics," IMA Journal of Applied Mathematics **25**, 241-265 (1980).
116. G. Nienhuis and L. Allen, "Paraxial wave optics and harmonic oscillators," Phys. Rev. A **48**, 656 (1993).
117. S. Yu, X. Piao, and N. Park, "Controlling random waves with digital building blocks based on supersymmetry," Phys. Rev. Appl. **8**, 054010 (2017).
118. R. G. Dorsch, A. W. Lohmann, Y. Bitran, D. Mendlovic, and H. M. Ozaktas, "Chirp filtering in the fractional Fourier domain," Applied optics **33**, 7599-7602 (1994).
119. N. M. Atakishiyev and K. B. Wolf, "Fractional fourier–kravchuk transform," JOSA A **14**, 1467-1477 (1997).
120. R. Gordon, "Harmonic oscillation in a spatially finite array waveguide," Opt. Lett. **29**, 2752-2754 (2004).
121. L. Verslegers, P. B. Catrysse, Z. Yu, and S. Fan, "Deep-subwavelength focusing and steering of light in an aperiodic metallic waveguide array," Phys. Rev. Lett. **103**, 033902 (2009).
122. D. N. Christodoulides, F. Lederer, and Y. Silberberg, "Discretizing light behaviour in linear and nonlinear waveguide lattices," Nature **424**, 817-823 (2003).
123. S. Weimann, A. Perez-Leija, M. Lebugle, R. Keil, M. Tichy, M. Gräfe, R. Heilmann, S. Nolte, H. Moya-Cessa, and G. Weihs, "Implementation of quantum and classical discrete fractional Fourier transforms," Nat. Commun. **7**, 11027 (2016).
124. M. Markowitz, K. Zelaya, and M.-A. Miri, "Auto-calibrating universal programmable photonic circuits: hardware error-correction and defect resilience," Opt. Express **31**, 37673-37682 (2023).
125. K. Zelaya, M. Markowitz, and M.-A. Miri, "The Goldilocks principle of learning unitaries by interlacing fixed operators with programmable phase shifters on a photonic chip," Sci. Rep. **14**, 10950 (2024).
126. K. Zelaya and M.-A. Miri, "Integrated photonic fractional convolution accelerator," Photonics Research **12**, 1828-1839 (2024).
127. M. Markowitz, K. Zelaya, and M.-A. Miri, "Learning arbitrary complex matrices by interlacing amplitude and phase masks with fixed unitary operations," Phys. Rev. A **110**, 033501 (2024).
128. K. Levenberg, "A method for the solution of certain non-linear problems in least squares," Quarterly of applied mathematics **2**, 164-168 (1944).
129. D. W. Marquardt, "An algorithm for least-squares estimation of nonlinear parameters," Journal of the society for Industrial and Applied Mathematics **11**, 431-441 (1963).